\documentstyle[12pt,fleqn]{article}

\newcommand{\sqrttriangle}{\sqrt{ -\triangle}}

\begin{document}

\hfill{Alberta-Thy-18-97}\\  

\hfill{hep-th/9709074} \\

\begin{center}
{\Large\bf Finite Temperature Nonlocal Effective Action for Scalar Fields}
\end{center}
\vspace{0.37truein}
\centerline{\bf Yuri Gusev\footnote{e-mail: ygusev@phys.ualberta.ca}}
\vspace{0.015truein}
\centerline{\em Theoretical Physics Institute, 
University of Alberta}
\centerline{\em Edmonton, Alberta, Canada T6G 2J1}
\vspace{10pt}
\centerline{\bf Andrei Zelnikov\footnote{e-mail: zelnikov@phys.ualberta.ca}}
\vspace*{0.015truein}
\centerline{\em Theoretical Physics Institute,  University of Alberta}
\centerline{\em Edmonton, Alberta, Canada T6G 2J1}
\centerline{and}
\centerline{\em P.N. Lebedev Physics Institute}
\centerline{\em Leninskii prospect 53, Moscow 117 924 Russia}
\vspace{10mm}

\centerline{\bf Abstract}  
           
\noindent
{\small	Scalar fields at finite temperature are considered 
in four dimensional ultrastatic curved spacetime.
One loop nonlocal effective action at finite temperature 
is found up to the second order in curvature expansion.
This action is explicitly infrared finite.
In the high temperature expansion of free energy,
essentially nonlocal terms linear in temperature are derived.}
\vfill
\thispagestyle{empty}
\pagebreak

\section{Introduction}

Effective action is a powerful tool in quantum  field theory. 
It contains information about all Green functions,
vacuum polarization effects, etc. In general case it is a non-local functional.
To study thermodynamic properties of a system 
one has to know the effective action at non-zero temperature. 
In this letter we calculate the one loop nonlocal effective action
and free energy for 
quantum scalar fields $\varphi$ in a curved spacetime 
at finite temperature $T$. At first we introduce
general notations for both zero and finite temperature theories
on some $D$-dimensional manifold.
We consider fields $\varphi$ that satisfy the equation 
	\begin{equation}
	{F}(\nabla) \varphi = 0, \ \ \
	{F}(\nabla) = \Box + P(x) - 
	{1\over 6} R(x).  \label{intro-2}
	\end{equation}
Here the Laplace-Beltrami operator
	$\Box \equiv g^{\mu\nu}\nabla_\mu\nabla_\nu$  
is constructed in terms of a covariant derivative $\nabla_{\mu}$. 
The metric $g^{\mu\nu}$ is
characterized by its Riemann and Ricci curvatures
	$R^\mu_{\,\cdot\, \alpha\nu\beta}=\partial_\nu
	\Gamma^\mu_{\alpha\beta}- \ldots,\ \ 
	R_{\alpha\beta}= R^\mu_{\,\cdot\, \alpha\mu\beta}$.\
The term with scalar curvature $R/6$ in the operator  ${F}(\nabla)$ 
is explicitly singled out for convenience only. The potential $P(x)$ has    
an  arbitrary structure and may include a curvature $R(x)$
or an interaction potential. This means that models satisfying 
eq.~(\ref{intro-2}) are quite general and 
describe  non-conformal and interacting scalar fields.

The one loop effective action 
can be expressed in terms of eigen values of the operator ${F}(\nabla)$ 
and formally reads
$	W=\frac12\,{\mathrm{Tr}}\, {\rm ln}\, F$. 
It can be written \cite{DeWitt-book65,BarVilk-PRep85}  
in the form of an  integral  over the proper time $s$,  
	\begin{equation}
	W=-\frac12\int^\infty_0 \frac{{\mathrm{d}}s}s {\mathrm{Tr}} K(s). \label{intro-7}
	\end{equation}
In coordinate representation ${\mathrm{Tr}}$ denotes the functional trace
$	{\mathrm{Tr}} \, K(s)=\int  {\mathrm{d}}^D x {K}(s|x,x)$
of the heat kernel, which is defined as a solution of the equations
	\begin{eqnarray}
	{{\mathrm{d}}\over {\mathrm{d}}s}{K}(s|x,y)=
F(\nabla^x) K(s|x,y), \hskip 2cm K(0|x,y)=\delta(x-y). \label{heateq}
	\end{eqnarray}
Knowledge of the heat kernel with separated points
allows one to construct the covariant diagrammatic expansion
of the effective action to all loop orders \cite{DeWitt-book65,BarVilk-PRep85}.
For our case of one  loop computations, it is sufficient to find the trace of the heat kernel.

Thermodynamics of a system is well defined if
background fields are static, so we restrict ourselves to considering
static spacetimes.  
It is well known \cite{Fradk59,MartSchwin59,DolJack-PRD74} that
a quantum field theory defined in a Euclidean spacetime, which is
periodic (for bosons) in imaginary time $\tau$ with a period $\beta$, 
amounts to a thermal field theory with temperature $T=1/\beta$.
In this letter we consider Euclidean manifolds
with the topology 
of a cylinder  $S^1\times R^3$.
The periodic (thermal) heat kernel ${K}^{\beta}$  can be expressed  as an  
infinite sum of zero temperature (vacuum) heat kernels
\cite{DolJack-PRD74,DowkCrit77}
	\begin{equation}
	{K}^{\beta}(s|\tau, \mbox{\boldmath $x$}; \tau', 
	\mbox{\boldmath $x$}')
	=\sum_{n=-\infty}^{\infty} 
	{K}(s|\tau, \mbox{\boldmath $x$}; {\tau}'+\beta n, 
	\mbox{\boldmath $x$}')         \label{sumK}
	\end{equation}
This  image sum  is equivalent to summation
over Matsubara frequencies in a momentum space representation
in thermal field theory.

We restrict now a class of spacetimes to the one possessing 
ultrastatic (optical) metrics      
	\begin{equation}
	{\mathrm{d}} s^2=g_{\mu\nu} {\mathrm{d}} x^\mu dx^\nu=
{\mathrm{d}} \tau^2+\bar{g}_{ij}(\mbox{\boldmath $x$})
 {\mathrm{d}} x^i  {\mathrm{d}} x^j      \label{us-metric}
	\end{equation}
where $\mu,\nu = (0,1,2,3)$ and $i,j = (1,2,3)$, 
and the metric tensor 
$\bar{g}_{ij}(\mbox{\boldmath $x$})$ depends on the spatial coordinates 
\mbox{\boldmath $x$} only. 
To  take into account temperature effects in curved spacetime explicitly 
one has to factorize the heat kernel into temporal and spatial $K^{(3)}(s)$ parts,
\begin{equation}
	\bar{K}(s|\tau, \mbox{\boldmath $x$}; \tau', 
	\mbox{\boldmath $x$}')
	= \frac{1}{(4\pi s)^{1/2}} \exp ({-\frac{(\tau - {\tau}')^2}{4s}})
	\bar{K}^{(3)}(s|\mbox{\boldmath $x$};  
	\mbox{\boldmath $x$}'),         \label{splitK}
\end{equation}
what is possible to do in ultrastatic spacetimes.
Here and below we denote objects in ultrastatic spacetimes by overbars.         
Then, the trace of  (\ref{sumK}) takes an elegant form 
\cite{DowkSch89,BarFrolZel95}  in terms of Riemann theta function, which is
defined in a usual way  $\theta_3 (a,b)= \sum_{n=-\infty}^{n=\infty}
{\mathrm{e}}^{2na{\mathrm{i}}} b^{n^2}$,
	\begin{equation}
	{\mathrm{Tr}}\bar{K}^{\beta}(s)= 
	\theta_3 
	\Big(0,
 	 {\mathrm{e}}^{-\frac{\beta^2}{4s}} 
	\Big) 
\frac{\beta}{(4\pi s)^{1/2}} 
\int {\mathrm{d}}^3 {\mbox{\boldmath $x$}} \, 
\bar{K}^{(3)}(s|\mbox{\boldmath $x$}, \mbox{\boldmath $x$}).     \label{thetaTrK}
	\end{equation} 

\section{Nonlocal effective action and free energy at finite temperature}

Various methods are applicable for calculation of ${\mathrm{Tr}}\bar{K}$.
A traditional tool of quantum field theory in curved spacetime,
the Schwinger-DeWitt series \cite{DeWitt-book65,BarVilk-PRep85},
corresponding to the heat kernel
expansion at small proper times $s$, 
is useful only for obtaining high temperature expansion
of effective action \cite{DowkSch89}.
We resort to a method of covariant 
perturbation theory \cite{CPT1,CPT2,CPT3,CPT4},
since local representations for the heat kernel
are not adequate to our problem.
This technique, as proposed by Vilkovisky in ref.~\cite{Vilk-Gospel},
is a regular procedure for covariant expansion of $K(s)$
in powers of curvatures (field strengths). 
This is essentially {\em nonlocal\/} representation,
and it takes into account an infinite number of 
derivatives acting on curvatures.
At small $s$ it reproduces the local Schwinger-DeWitt expansion 
\cite{CPT2,BGVZ-JMP94-asymp}.

Besides,  the Schwinger-DeWitt expansion,
which is very effective in dealing with ultraviolet divergencies
of the effective action \cite{BarVilk-PRep85}, 
fails in the infrared (large $s$) regime for massless theories.
In contrast, the one loop effective action and  all Green functions
computed in the framework of covariant perturbation theory
are infrared finite if the dimension of
a manifold under consideration is larger than two
\cite{CPT1,CPT2,BGVZ-NPB95,BGVZ-JMP94-asymp}.
In this letter we show that the one loop nonlocal  effective action 
for arbitrary scalar  fields calculated on a four dimensional manifold 
remains infrared finite at finite temperature. 
This fact is very important, for infrared divergencies are
one of major obstacles in  thermal field theory \cite{DrumHorLandReb-PLB97}.

It is necessary now to give a brief description of covariant perturbation
theory at zero temperature in arbitrary dimensions $D$. 
Let us first display a structure of the nonlocal  trace of the heat kernel \cite{CPT2},
	\begin{eqnarray}
	{\mathrm{Tr}} K(s) &=& \frac1{(4\pi s)^{D/2}}\int\! {\mathrm{d}}^{D}x\,
	g^{1/2}\, \Big\{1+s{P}
	+s^2 \Big[
	R_{\mu\nu} f_{1}(-s\Box) R^{\mu\nu}
	\nonumber\\[2mm]&& \mbox{}
	+ R f_{2}(-s\Box) R 
	+ {P} f_{3}(-s\Box) R
	+ {P}  f_{4}(-s\Box) {P}
	 \Big]\Big\}
	+{\mathrm{O}}[\Re^3].  \label{TrK}
	\end{eqnarray}
Form factors $f_i $ are analytic functions
of the dimensionless argument 
	$s\Box$.
They act on tensor
invariants constructed of
the set of field strengths   $R^{\alpha\beta},\  P$ (called here curvatures)
characterizing  background. 
We use for these curvatures the collective notation $\Re$. 
First two terms of the sum (\ref{TrK}) are purely local and coincide
with first two coefficients of the short time expansion \cite{CPT2}.
The Euclidean spacetime is supposed to be  asymptotically flat 
and  have the topology $R^{D}$.
Therefore, gravitational curvatures 
and potential ${P}$ should vanish at infinity \cite{CPT2}.
The calculations in covariant perturbation
theory are carried out with accuracy ${\mathrm{O}}[\Re^n]$, i.e.,
up to terms of $n$-th and higher power in the curvature $\Re$.
Thus, this curvature expansion is valid for highly inhomogeneous
background fields, $\nabla \nabla \Re >> \Re^2$.

All form factors in (\ref{TrK}) can be  
expressed in terms of one  basic form factor
	\begin{equation}
	f(-s\Box)=\int_0^1\!\! {\mathrm{d}}\alpha\: 
{\mathrm{e}}^{\alpha(1-\alpha)s\Box}. \label{basicf}
	\end{equation} 
Their explicit form reads \cite{CPT2}
	\begin{eqnarray}
	f_1(-s\Box) &=& \frac{f(-s\Box)-1-\frac16 s\Box}{(s\Box)^2},   \nonumber
	\\
	f_2(-s\Box) &=& \frac18\left[
  	\frac1{36}\,f( -s \Box )-\frac13\,\frac{f( -s \Box )-1}{ s \Box }-
  	\frac{f( -s \Box )-1 - \frac16 s \Box }{ (s \Box)^2}
  	\right],  \nonumber
	\\
	f_3( -s \Box ) &=&
  	\frac1{12}\,f( -s \Box )-\frac12\,\frac{f( -s \Box )-1}{ s \Box },  
	\\     \label{formfactors}
	f_4( -s \Box ) &=& \frac12 \,f( -s \Box ).  \nonumber 
	\end{eqnarray}

Let us return now to the finite temperature case. 
The free energy for scalar fields in ultrastatic spacetime
$\bar{F}^{\beta}$ is defined in terms of the Euclidean
effective action at finite temperature  $\bar{W}^{\beta}$ by the relation
\begin{equation}
\bar{F}^{\beta} \equiv 
\frac1{\beta} \bar{W}^{\beta}=
\mbox{}-\frac1{2\beta} 
\int_0^{\infty} \frac{\mathrm{d}s}{s} 
{\mathrm{Tr}} \bar{K}^{\beta} (s).
\end{equation}
It is convenient to separate a vacuum mode $~n=0~$ 
from the infinite sum in (\ref{sumK}), since 
only vacuum effective action contains ultraviolet divergencies \cite{DowkKenn78}.
This is equivalent to subtracting  the zero temperature free energy  
$\bar{F}^{\infty}$ from $\bar{F}^{\beta}$.
Since only renormalized effective action and free energy have physical meaning,
we  renormalize one loop vacuum effective action (\ref{intro-7})
with the  use of  zeta function regularization: 
	\begin{equation}
	\bar{W}^{\infty}_{\mathrm{ren}}=
	\mbox{}-\frac12 \frac{\partial}{\partial\epsilon}\left[
	\frac{\mu^{2\epsilon}}{\Gamma(\epsilon)}
	\int_0^{\infty}\! \frac{{\mathrm{d}}s}{s^{1-\epsilon}}\, 
	{\mathrm{Tr}} \bar{K}^{\infty}(s) 
	\right]_{\epsilon=0},         \label{zeta-reg}
	\end{equation}
where $\mu$ is a mass-like regularization parameter and $\Gamma$ is
the gamma function.                          
$\bar{F}_{\mathrm{ren}}^{\infty}$  will  be combined with $n \neq 0$ 
terms at the end of our derivations.
Thus, we compute
\begin{equation}
	\bar{F}^{\beta}_{\mathrm{ren}} - \bar{F}^{\infty}_{\mathrm{ren}} =
\mbox{}-\frac{1}{2}   
\int_0^{\infty}\! \frac{{\mathrm{d}} s }{s}
\Big(    \Theta_3 
	\big(0,
 	 {\mathrm{e}}^{-\frac{\beta^2}{4s}} 
	\big)                                             
	-1   \Big)       
\frac{1}{(4 \pi s)^{1/2}}
\int {\mathrm{d}}^3 x \, 
\bar{K}^{(3)} (s|\mbox{\boldmath $x$}; \mbox{\boldmath $x$}).     \label{Wb-W}
\end{equation}                
The three dimensional heat kernel $\bar{K}^{(3)}(s)$ is defined as a solution of (\ref{heateq})
with the three dimensional operator (\ref{intro-2}). In this case the operator 
$\Box$ is the three dimensional Laplacian $\triangle$, and the curvatures 
$\bar{\Re}=(\bar{P}(\mbox{\boldmath $x$}), \bar{R}_{ij}(\mbox{\boldmath $x$}))$ 
are defined on a three dimensional hypersurface $\tau={\mathrm{const}}$. 

The integral (\ref{Wb-W}) with first two terms  of 
the trace of the heat kernel (\ref{TrK})
gives just numerical coefficients.
This local contribution  to free energy is known 
 \cite{BarFrolZel95,DowkKenn78}
and coincides with first two terms of high temperature expansion  \cite{DowkSch88}.
The problem reduces to computation of the thermal form factors 
\begin{equation}
\gamma^{\beta}_{i}(-\triangle) =
\int_0^\infty {{\mathrm{d}}y\over y}
\left[\theta_3\Big(0,{\mathrm{e}}^{-y}\Big)-1 \right] 
f_{i}   \Big(  -\frac{\beta^2}{4 y}  \triangle  \Big),            \label{gammaTi}
\end{equation}
where  $y=\beta^2/4s$.                                                                                  
                                           
Calculation of (\ref{gammaTi}) is a key technical  part of this letter and we
explain its idea on the example of the basic form factor (\ref{basicf}). 
Straightforward integration over $y$ leaves us with the 
modified Bessel function of the second kind
\begin{eqnarray}
\gamma^{\beta}(-\triangle)&=& 
4 \sum_{n=1}^\infty 
\int_0^1 {\mathrm{d}}\alpha~
K_0\left(n\beta\sqrt{\alpha(1-\alpha)}\sqrttriangle\right).  \label{Bessel0}
\end{eqnarray} 
Another change of variables, $x = 2 \sqrt{\alpha(1-\alpha)}  $,  
allows us to find its form in terms of the exponential integrals, 
\begin{equation}
\int_0^1 {\mathrm{d}}x~{x\over\sqrt{1-x^2}}~K_0(\frac{nz~x}{2})
={1\over nz}\Big[\,   {\mathrm{Ei}}\big(\frac{nz}{2}\big)   {\mathrm{e}}^{-{nz}/{2}}
-   {\mathrm{Ei}}   \big(-\frac{nz}{2}\big) {\mathrm{e}}^{{nz}/{2}}\, \Big],         \label{Ei}
\end{equation}  
where $z= \beta\sqrttriangle$.  
Now we can use for the right hand side of (\ref{Ei}) 
its standard form in terms of elementary functions and obtain
\begin{equation}
\gamma^{\beta}(-\triangle)=
4 \int_0^\infty \! {\mathrm{d}}t \, \sum_{n=1}^\infty 
\,{\sin(t)\over t^2+(nz)^2/4}.  \label{sinus}
\end{equation}  
This is just what we need since the sum over $n$ can be exactly evaluated,
\begin{equation}
\sum_{n=1}^{\infty} \frac{1}{t^2+n^2 z^2 /4}=
\frac12 \left[\frac{2\pi}{zt}\,\frac{1}{{\rm th}(2\pi t/z)}
-\frac{1}{t^2}\right].
\end{equation}
There are, as seen from (\ref{formfactors}), two other types of 
basic thermal form factors  with one and two subtractions. 
Derivations for them are  similar but more technically involved.

Finally, we arrive at the following integral representation,
\begin{eqnarray}
\bar{F}^{\beta}_{\mathrm{ren}} - \bar{F}^{\infty}_{\mathrm{ren}}&=&
\mbox{}- \frac1{32\pi^2}
\int \!  {\mathrm{d}}^3 x \, \bar{g}^{1/2}\, 
        \Big\{
\frac{16}{45}\frac{\pi^4}{\beta^4}+
\frac{4}{3} \frac{\pi^2}{\beta^2} {\bar{P}}
\nonumber\\[2mm]&&\mbox{}
+ \Big[\,
\bar{R}_{i j} \gamma_1^{\beta}(-\triangle) \bar{R}^{i j}
+ \bar{R} \gamma_2^{\beta}(-\triangle) \bar{R} 
\nonumber\\[2mm]&&\mbox{} 
+ \bar{P} \gamma_3^{\beta}(-\triangle) \bar{R} 
+ {\bar{P}} \gamma_4^{\beta}(-\triangle) {\bar{P}}   \,
\Big]
        +{\mathrm{O}}[\bar{\Re}^3]\Big\}.  \label{efacT}
        \end{eqnarray}
Here the thermal form factors have a form
\begin{equation}       
 \gamma_i^{\beta}(-\triangle) = \int_0^\infty\!\! {\mathrm{d}}t \,\, g_i(t)
\left[ 
\frac{2\pi}{ z t}\, \frac{1}{{\rm th}(2\pi t/z)} 
- \frac{1}{t^2}    \label{gammaTgen}
\right],   
\end{equation}
and $g_i \ (i=1,...4)$\ are the simple combinations of elementary functions
\begin{eqnarray}       
 g_1(t)&=&  
 \mbox{}-\frac12 \Big( \frac{\sin (t)}{t^2} 
+ 3\Big[{\cos (t)\over t^3}-{\sin(t)\over t^4}\Big]\Big),
\nonumber\\ [2mm]
 g_2(t)&=& \frac1{48}\Big(
  \frac1{3} \sin (t) 
+ 2 {\cos (t)\over t}
+ {\sin(t)\over t^2}
+ 9\Big[ {\cos (t)\over t^3}
	- \frac{\sin(t)}{t^4}\Big]\Big),
\nonumber\\   [2mm]
  g_3(t)&=& \frac12\Big(
  \frac1{3} \sin (t) 
+{\cos (t)\over t}
- {\sin(t)\over t^2}\Big),  \label{gammasT}  
\\           [2mm]
   g_4(t)&=& \sin(t). \nonumber 
\end{eqnarray}

We have achieved our goal of evaluating the image sum (\ref{sumK}),
moreover the answer is expressed in terms of elementary functions.
As can be seen from eq.~(\ref{gammaTgen}) 
'a golden rule' of field theory calculations \cite{Vilk-Gospel},
namely, to keep a proper time integration to the very end,  
has been also satisfied, as the variable $t$ (which is dimensionless in contrast to
the proper time $s$) clearly plays a role of  the proper time. 
One can check that integrals of this type are regular,
and two apparent singularities at the lower limit
compensate each other.
The price for all this is that an argument of the hyperbolic tangent function is 
the dimensionless combination $\beta\sqrttriangle$.

The final result for renormalized free energy at finite temperature   
$\bar{F}_{\mathrm{ren}}^{\beta}$
is presented by a sum of   eqs. (\ref{efacT})--(\ref{gammasT}) and renormalized
free energy at zero temperature $\bar{F}^{\infty}_{\rm ren}$.
After the  zeta regularization (\ref{zeta-reg}), the latter one takes the  form,
	\begin{eqnarray}
        \bar{F}^{\infty}_{\mathrm{ren}}&=&\mbox{}
	-\frac1{32\pi^2}\int\! 
	{\mathrm{d}}^3 x\, \bar{g}^{1/2}\, 
        \Big\{ \bar{R}_{i j } \gamma_1(-\triangle) \bar{R}^{i j }
	+ \bar{R} \gamma_2(-\triangle) \bar{R} \nonumber\\[2mm]&& \mbox{}
	+ \bar{P} \gamma_3(-\triangle) \bar{R}
	+ \bar{P} \gamma_4(-\triangle) \bar{P}   
        +{\mathrm{O}}[\Re^3]\Big\}.       \label{efacz}
        \end{eqnarray}
where form factors $\gamma_i(-\triangle)$,\ $i=1, ... 4$, are
	\begin{eqnarray}
	\gamma_1(-\triangle)&=&~\frac1{60} \left[-{\rm ln}\Big(-\frac{\triangle}{\mu^2}\Big)
	+\frac{46}{15}\right],~~~ 
\gamma_3(-\triangle)=-\frac1{18},  \nonumber\\
\gamma_2(-\triangle)&=&\frac1{180}\left[
{\rm ln}\Big(-\frac{\triangle}{\mu^2}\Big)-\frac{97}{30}\right],~~~
\gamma_4(-\triangle)=\frac12 \left[-{\rm ln}\Big(-\frac{\triangle}{\mu^2}\Big)+ 2 \right]. 
\label{gammas}
\end{eqnarray}


\section{High temperature expansion of free energy}

Let us emphasize that the formulae ({\ref{efacT})--(\ref{gammasT}) 
are valid at arbitrary finite temperature. 
Unfortunately direct evaluation of integrals (\ref{gammaTgen}) is possible only
if one studies asymptotic behavior of the free energy, e.g., 
high temperature regime. 
It can be done relatively easy, because, 
as usual, the problem boils down to finding 
 $\beta \rightarrow 0$ asymptotic of the
thermal form factors (\ref{gammaTgen})--(\ref{gammasT}). 
One has to be careful while dealing with mutually compensating
singularities in $\gamma_{i}^{\beta}$. After some algebra, 
the outcome for, e.g. (\ref{basicf}), is, 
\begin{eqnarray}
{\gamma}^{\beta}  
(-\triangle) &=&
\frac{2\pi^2}{\beta \sqrttriangle }
+2\Big[
	{\rm ln}(\frac{\beta\sqrttriangle}{4 \pi})-1 +{\mathrm{C}}
\Big]
\nonumber\\&&\mbox{}
-\frac{\zeta(3)}{24\pi^2}\beta^2 (-\triangle) 
+\frac{\zeta(5)}{640\pi^4}\beta^4 (-\triangle)^2 + {\mathrm{O}}[\beta^6],
\ \ \ \beta \rightarrow 0,                             \label{basicgammaT}
\end{eqnarray}                  
where ${\mathrm{C}}$ is the Euler constant and $\zeta$ is the Riemann zeta function.

Having  (\ref{efacT}) expanded in temperature series,
we can add the vacuum free energy (\ref{efacz}) to it. 
The result for the high temperature expansion
of the renormalized one loop free energy takes a form,
\begin{eqnarray}
\bar{F}_{\mathrm{ren}}^{\beta}&=&
\mbox{}- \int\! {\mathrm{d}}^3 x
\, \bar{g}^{1/2}\, 
        \Big\{
\frac{\pi^2}{90\beta^4}
+ \frac{1}{24\beta^2} {\bar{P}}
\nonumber\\[2mm]&&\mbox{}
+ \frac{1}{32\beta}    
\Big[
\frac{1}{16}
	\bar{R}_{i j} 
\frac{1}{\sqrttriangle}
	\bar{R}^{i j }
-\frac{25}{1152} 
	\bar{R} 
\frac{1}{\sqrttriangle}
	 \bar{R} 
\nonumber\\[2mm]&& \mbox{}
-\frac{1}{12} 	
	{\bar{P}} 
\frac{1}{\sqrttriangle}
	\bar{R}
+ 	{\bar{P}}
\frac{1}{\sqrttriangle}
	{\bar{P}}
\Big]  
\nonumber\\[2mm]&& \mbox{}
 +\frac{1}{16\pi^2} \Big(  
	{\rm ln}(\frac{\beta\mu}{4\pi}) +  {\mathrm{C}}
\Big)   \,
\Big[
	\frac{1}{60}  \bar{R}_{\mu\nu} \bar{R}^{\mu\nu}
	-\frac{1}{180} \bar{R}  \bar{R} 
	+ \frac12 {\bar{P}} {\bar{P}}
\Big]\Big\}
\nonumber\\[2mm]&& \mbox{}
        +{\mathrm{O}}[\bar{\Re}^3]
+ {\mathrm{O}}[\beta^2],
 \ \ \ \beta \rightarrow 0. \label{efachighT}
        \end{eqnarray}                      

First and foremost, one should note that $1/\beta$ order
has now form  $\Re \frac{1}{\sqrt{-\triangle}}\Re$. Therefore it is essentially
nonlocal, and techniques based on local short time expansions can not
generate it.  
We expect that terms of higher orders in curvatures 
\cite{CPT4,BGVZ-JMP94-asymp}  will also
give nonlocal contribution to the linear $T$ order.  
The meaning  of these structures
can be understood from an elegant spectral representation
\cite{CPT3,CPT4}  in terms of massive Green functions,
\begin{equation}
\frac{1}{\sqrttriangle}=
\frac{2}{\pi}\int_{0}^{\infty}\! {\mathrm{d}}m \, 
\frac{1}{m^2-\triangle}.  \label{spectral}
\end{equation}

A remarkable property of the expression (\ref{efachighT}) 
is that all $\ln (-\triangle)$ have disappeared, leaving 
logarithmic temperature dependance in the form of ${\rm ln}(\beta \mu)$,
the  combination known   in flat space thermal field theory  
\cite{DolJack-PRD74,HabWeld-PRD82,BranFrenk-hepth97}.

Combination of quadratic in curvatures terms at the logarithm
is nothing but the trace of the second Schwinger-DeWitt coefficient $a_2$,
taken with Riemann curvature expressed via Ricci one 
\cite{DeWitt-book65,BarVilk-PRep85,CPT2}. 
Higher powers of $\beta$ (see eq. (\ref{basicgammaT})), which we 
discarded in $\bar{F}_{\mathrm{ren}}^{\beta}$, 
are also quadratic in 
curvatures parts of higher Schwinger-DeWitt coefficients
\cite{BGVZ-JMP94-asymp}.
          
All local terms in our result (\ref{efachighT}) perfectly 
reproduce those of   ref.~\cite{DowkSch88}.
We also obtained the explicit form of all nonlocal terms
of the second order in curvatures. They contain form factor
     $1/\sqrttriangle$
and turn out to be proportional to $1/\beta$.
Implicitly these nonlocal parts are contained  in
the term denoted by $\bar{\zeta}'_{(3)} (0)$ in eq.~(17) of \cite{DowkSch88}.


\section{Conclusion}

The main result of this paper is
the  one loop nonlocal free energy for scalar fields 
at finite temperature, eqs.~(\ref{efacT})--(\ref{gammas}).  
Since in the setting of our problem $\bar{F}_{\beta}$ and $\bar{W}^{\beta}$
differ only by a factor $\beta^{-1}$, one can speak about free energy and effective action
interchangeably. One can derive other quantities, 
such as Green function, stress tensor or entropy, 
by varying this free energy  correspondingly with respect
to the potential, metric or temperature.
Clearly, to proceed to physical applications one first 
needs some kind of spectral representation for finite temperature form factors,
like the ones derived for  zero temperature effective action 
\cite{CPT1,CPT2,CPT3,CPT4}
or its high temperature limit (\ref{spectral}).   

It should be emphasized that the perturbation theory we used to derive 
our results works well for highly inhomogeneous background fields, i.e.,
$\nabla \nabla \Re >> \Re^2$.
Therefore, eqs. (\ref{efacT})--(\ref{gammas}), (\ref{efachighT})
are not applicable to the opposite case of constant or nearly constant 
background.

Our work also demonstrates, that infrared divergencies in finite temperature
field theory are artificial and a correctly computed thermal  effective action 
is infrared finite at all stages of calculations,
what supports the main idea of ref.~\cite{DrumHorLandReb-PLB97}.

Generalization of our derivations to arbitrary static spaces
does not seem to pose serious complications, and 
involves well elaborated techniques of conformal  transformations
to ultrastatic spacetimes  \cite{Dowk86,DowkSch88}.
In order to include into consideration manifolds with boundaries,
a major revision of the perturbation theory \cite{CPT2} is needed,
though it is still feasible to do.  
It is also possible to  include the chemical potential \cite{DowkSch89}.
Non-Abelian gauge fields can be taken into consideration as well,  but
the factorization formula (\ref{thetaTrK}) 
will be modified by additional terms 
\cite{LeonZeln-PLB92}, connected to temporal components of gauge fields.
One can also easily generalize the results to quantum fields of other spins.

\section*{Acknowledgments}

This work was partially supported by
National Science and Engineering Research Council
of Canada. Yu. G. is supported also
by National Research Fellowships from Canadian Institute for Theoretical
Astrophysics, and A. Z. is grateful to the Killam Trust for its financial support.
\pagebreak

\end{document}